\documentclass{elsart}
\usepackage{epsfig}

\newcommand{\eff}{\varepsilon}
\newcommand{\psip}{\psi(2S)}

\newcommand{\jpsi}{J/\psi}

\newcommand{\pip}{\pi^+}
\newcommand{\pin}{\pi^-}

\newcommand{\etac}{\eta_{c}}
\newcommand{\etackk}{\kp\kn\pip\pin\pip\pin}

\newcommand{\beq}{\begin{equation}}
\newcommand{\eeq}{\end{equation}}
\newcommand{\beqn}{\begin{eqnarray}}
\newcommand{\eeqn}{\end{eqnarray}}
\newcommand{\beqns}{\begin{eqnarray*}}
\newcommand{\eeqns}{\end{eqnarray*}}
\newcommand{\bfg}{\begin{figure}}
\newcommand{\efg}{\end{figure}}
\newcommand{\bitm}{\begin{itemize}}
\newcommand{\eitm}{\end{itemize}}
\newcommand{\bnum}{\begin{enumerate}}
\newcommand{\enum}{\end{enumerate}}
\newcommand{\btbl}{\begin{table}}
\newcommand{\etbl}{\end{table}}
\newcommand{\btbu}{\begin{tabular}}
\newcommand{\etbu}{\end{tabular}}

\newcommand{\kp}{K^+}
\newcommand{\kn}{K^-}
\newcommand{\ks}{K^{0}_{S}}

\newcommand{\rhoo}{\rho^0}
\newcommand{\g}{\gamma}
\newcommand{\be}{\begin{enumerate}}
\newcommand{\ee}{\end{enumerate}}
\newcommand{\bi}{\begin{itemize}}
\newcommand{\ei}{\end{itemize}}

\newcommand{\pio}{\pi^0}

\newcommand{\ar}{\rightarrow}

\renewcommand{\thefootnote}{\fnsymbol{footnote}}

\begin{document}

\begin{frontmatter}
\title{\boldmath  First Measurements of $\etac$ Decaying into 
${\kp\kn}2({\pi^+\pi^-})$ and $3({\pi^+\pi^-})$}

\date{15 Jan. 2004}

\maketitle

%\begin{center}
\begin{center}
M.~Ablikim$^{1}$,    J.~Z.~Bai$^{1}$,               Y.~Ban$^{11}$,
J.~G.~Bian$^{1}$,    X.~Cai$^{1}$,                  J.~F.~Chang$^{1}$,
H.~F.~Chen$^{17}$,   H.~S.~Chen$^{1}$,              H.~X.~Chen$^{1}$,
J.~C.~Chen$^{1}$
% ...
\\(BES Collaboration)\\
\vspace{0.2cm}
%\label{att}
$^{1}${\it Institute of High Energy Physics, Beijing 100049, People's
Republic of
 China }\\
% ...
$^{11}$ {\it Peking University, Beijing 100871, People's Republic of
China}\\
% ...
$^{17}$ {\it University of Science and Technology of China, Hefei 230026,
People's Republic of
China}\\
% ...

\end{center}
\vskip 0.3cm

%\end{center}
%\date{}
%\clearpapge
%\begin{document}

\begin{abstract}
The decays of $\eta_c$ to $\kp\kn 2(\pip\pin)$ and $3(\pip\pin)$ are
observed for the first time
using a sample of $5.8\times 10^7$ $\jpsi$ events collected by the BESII
detector. The product branching fractions are determined to be 
$B(\jpsi\ar\g\etac)\cdot B(\etac\ar\kp\kn\pip\pin\pip\pin)$ $= (1.21 \pm 0.32\pm
0.23)\times 10^{-4}$,   
$B(\jpsi\ar\g\etac)\cdot B(\etac\ar  K^{*0}\overline{K}^{*0}\pip\pin)=
(1.29\pm 0.43\pm 0.32)\times 10^{-4}$, and  $B(\jpsi\ar\g\etac)\cdot 
B(\etac\ar\pip\pin\pip\pin\pip\pin)=
(2.59 \pm 0.32\pm0.48)\times 10^{-4}$ . The upper limit 
for $\etac\ar\phi
\pip\pin\pip\pin$   is also obtained as 
$B(\jpsi\ar\g\etac)\cdot 
B(\etac\ar\phi\pip\pin\pip\pin)< 6.03\times 10^{-5} $ at the 90\% confidence level.

\vspace{3\parskip}

\noindent{\it PACS:} 13.25.Gv, 14.40.Gx, 13.40.Hq

\end{abstract}
\end{frontmatter}
%\clearpage

\section{Introduction}   \label{introd}

The $\eta_c$, a $~^1S_0$ state in the charmonium family, was found in
the inclusive photon spectra from $\jpsi$ and $\psip$~\cite{cryball}
decays, as well as in hadronic decays~\cite{markiietac}.  A number of
decay modes of $\eta_c$ were then measured~\cite{markdm2}. More recent
measurements of hadronic decays of $\etac$ are listed in
Ref.~\cite{besiibelle}. According to Ref.~\cite{quigg}, the $\etac$ is
expected to have numerous decay modes into hadronic final states.
Although a number of decay modes of $\etac$ have been measured by
different experimental collaborations, the number of measured $\eta_c$
decay channels are few. This means that many decay modes of $\etac$
are unknown. The 58 million, $(57.70 \pm 2.72) \times
10^6$~\cite{fangss}, $\jpsi$ events taken at BESII provide a chance to
observe new decays. In this analysis,
$\etac$ decaying into $\kp\kn \pip\pin\pip\pin$ and
${\pi^+\pi^-\pi^+\pi^-\pi^+\pi^-}$ are studied using
$\jpsi\ar\g\etac$.
% with $5.8 \times 10^7$ BESII $\jpsi$ events.

The upgraded Beijing Spectrometer detector, located
at the Beijing Electron-Positron Collider (BEPC),
is a large solid-angle
magnetic spectrometer which is described in detail in Ref.~\cite{besii}.
The momentum of the charged particle is determined by a
40-layer cylindrical main drift chamber (MDC) which has a momentum
resolution of
 $\sigma_{p}$/p=$1.78\%\sqrt{1+p^2}$ ($p$ in GeV/c).
Particle identification is accomplished by specific ionization ($dE/dx$)
measurements in the drift chamber and time-of-flight (TOF) information in
a barrel-like array of 48 scintillation counters. The $dE/dx$ resolution
is $\sigma_{dE/dx}=8.0\%$; the TOF resolution for Bhabha events is
$\sigma_{TOF}=180$ ps.
Radially outside of the time-of-flight counters is a 12-radiation-length
barrel shower counter (BSC) comprised of gas
tubes interleaved with lead sheets. The BSC measures
the energy and direction of photons with resolutions
of $\sigma_{E}/E\simeq21\%\sqrt{E}$ ($E$ in GeV), $\sigma_{\phi}=7.9$ mrad, and
$\sigma_{z}=2.3$ cm. The iron flux return of the magnet is instrumentd
with three double layers of counters that are used to identify muons.

A GEANT3 based Monte Carlo package (SIMBES) with detailed
consideration of the detector performance
is used to obtain the detection efficiency. The consistency between data 
and Monte Carlo has been carefully
checked in many high purity physics channels, and the agreement is
reasonable~\cite{simbes}. 

\section{\boldmath Analysis of $\jpsi\ar\g\etac, ~\etac\ar\kp\kn\pip\pin\pip\pin$}
These events are observed in the topology $\g\kp\kn\pip\pin\pin\pin$.
Events with six good charged tracks and at least one isolated photon
are selected. 
The  selection criteria for good charged tracks and isolated photons
are described in detail in Ref.~\cite{rhopi}. Each charged track must be
well fitted to a helix, originating from the interaction region of R$_{xy}$
$<$ 2 cm and $|z|~<$ 20 cm, and have a polar angle $\theta$ in the range
$|\cos\theta|$ $<$ 0.8. Here  R$_{xy}$ is the distance from the beam
axis, and $z$ is along the beam axis.  Isolated photons
are those that have energy deposited in the BSC greater than 60 MeV, the
angle between the direction at the first layer of the BSC and the developing
direction of the cluster less than 30$^{\circ}$, and the angle between
photons and any charged tracks larger than $5^{\circ}$. Two of the charged
tracks should be identified as kaons by combined TOF and dE/dx information.

A four-constraint (4C) kinematic fit is performed under the hypothesis of
$J/\psi \to \g\kp\kn\pip\pin\pip\pin$,
and the $\chi^{2}_{\g\kp\kn\pip\pin\pip\pin}$ is
required to be less than 10.
To reject background from $\jpsi\ar\g\g\kp\kn\pip\pin\pip\pin$,
$\chi^{2}_{\g\etackk}$ is required to be less than
$\chi^{2}_{\g\g\etackk}$. Background events from $\jpsi\ar\etackk$
are eliminated by requiring $\chi^{2}_{\g\etackk}<$ $\chi^{2}_{\etackk}$  
and $P_{miss} > 55$ MeV/c, where $P_{miss}$
is the missing momentum of charged tracks.

\begin{figure}[htpb]
\centerline{\includegraphics[width=0.5\textwidth,height=0.32\textheight]
        {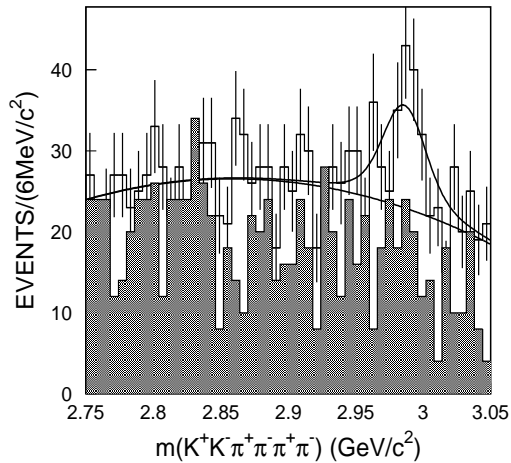}}
\caption{The distribution of $m_{\etackk}$ for selected events. The histogram with error bars
is from data, the shaded part is the background estimated from 
$J/\psi \to anything$ Monte Carlo simulation, and the curve represents the
fitting results described in the text.}
\label{metackk}

\end{figure}

After the above selection, the $K^+ K^- \pi^+ \pi^- \pi^+ \pi^-$
invariant mass, $m_{\etackk}$, distribution is shown in Fig.
\ref{metackk}.  A peak at the $\etac$ mass is observed.  The shaded
histogram is the background estimated from 58 million $\jpsi\ar anything$
Monte-Carlo events generated with the Lund-charm
generator~\cite{chenjc}; no prominent signal in the $\etac$ mass
region is seen. Also, 100,000 events for the two possible background
channels $\jpsi\ar \kp\kn2(\pip\pin)$ and $\jpsi\ar\g 3(\pip\pin)$
are simulated. After final selection, no events remain in the $\etac$
mass region.  A Breit-Wigner folded with a Gaussian to take into
account the mass resolution of 12.3 MeV/c$^2$ at the $\eta_c$ and a
polynomial background are used in the fit.  The fit gives $100\pm 26$
$\eta_c$ events with a statistical significance of 4.0 $\sigma$, where
the mass and width of $\eta_c$ are fixed to the PDG values~\cite{pdg2004}.

\begin{figure}[htpb]
\centerline{\includegraphics[width=0.5\textwidth,height=0.32\textheight]
        {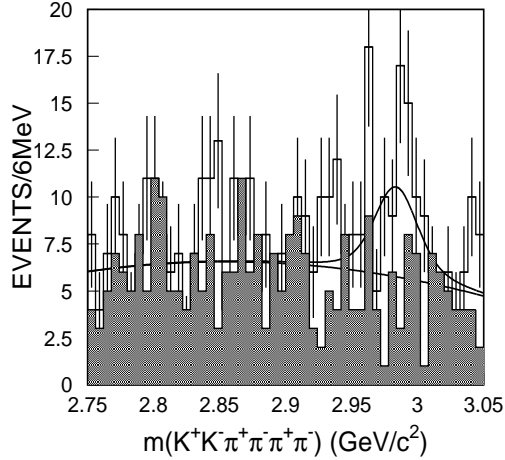}}
\caption{The distribution of $m_{\etackk}$ for $\etac\ar
K^{*0}\overline{K}^{*0}\pip\pin$ candidate events. The histogram with error bars
is for data, the shaded part is the background estimated from 
$K^{*0}$($\overline{K}^{*0}$) sidebands, and the curve is the fitting results
described in the text.}
\label{metac2kst}

\end{figure}

Using this sample, we search for the decay mode $\etac\ar
K^{*0}\overline{K}^{*0}\pip\pin$. To select $K^{*0}\overline{K}^{*0}\pip\pin$
events, we require that the invariant masses of $\kp\pin$ and $\kn\pip$
must statisfy $|m_{K\pi}-0.896|<0.05$ GeV/$c^2$.
After the $K^{*0}$ and
$\overline{K}^{*0}$ selection, the $\kp\kn 2(\pip\pin)$ invariant mass is
shown in Fig.~\ref{metac2kst}. A small peak at the $\etac$ mass is
observed.  The background events corresponding to the shaded histogram in
Fig.~\ref{metac2kst} are estimated from $K^{*0}$ and $\overline{K}^{*0}$
sidebands (0.1 GeV/c$^2$ $<|m_{\kp\pin}-0.896|<0.15$ GeV/$^/c^2$ and
0.1 GeV/c$^2$ $<|m_{\kn\pip}-0.896|<0.15$ GeV/$^/c^2$), and there is
no evident $\etac$ signal. $45\pm 15$ events are obtained by fitting
the mass spectrum with a Breit-Wigner folded with a Gaussian to
account for the $\etac$ mass resolution plus a second polynomial
background. The corresponding mass and width of the $\eta_c$ are fixed
to PDG values~\cite{pdg2004}.  Since the significance of the peak is
only 3$\sigma$, we also give the upper limit for $\etac\ar
K^{*0}\overline{K}^{*0}\pip\pin$. With the Bayes method, the fit of this
distribution yields 65 events at the 90\% confidence level.

\begin{figure}
%[htpb]
\centerline{\includegraphics[width=0.5\textwidth,height=0.32\textheight]
        {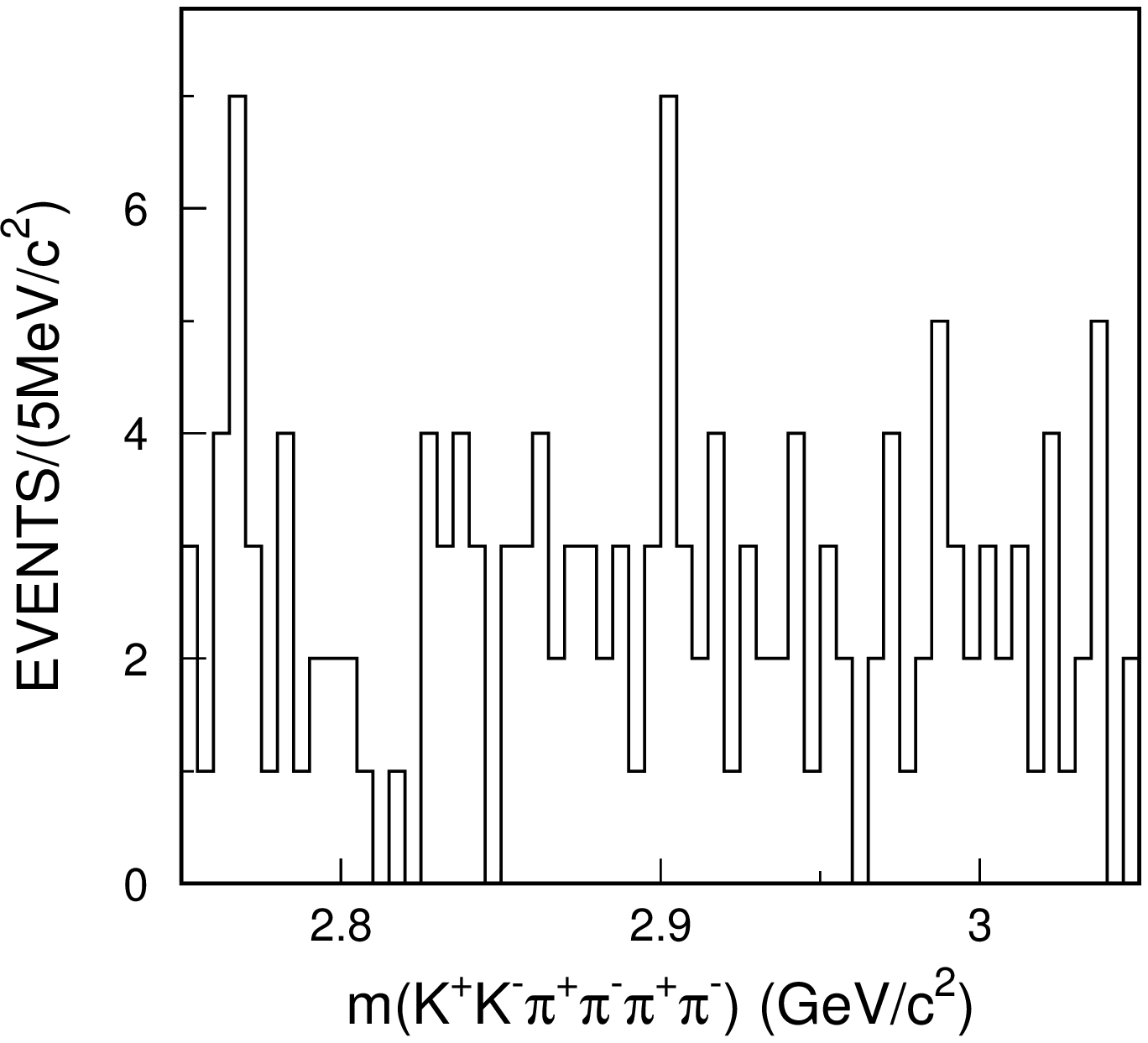}}
\caption{The distribution of $m_{\etackk}$ for $\etac\ar
        \phi\pip\pin\pip\pin$ candidate events.}

\label{phi4pi}
\end{figure}

The $\jpsi \ar \g\kp\kn\pip\pin\pip\pin$ sample can also be used to
search for $\etac\ar \phi\pip\pin\pip\pin$.  For selecting a $\phi$
signal, the $\kp\kn$ mass, $m_{\kp\kn}$, is required to be in the region
$|m_{\kp\kn}-1.02|<0.015$ GeV/c$^2$.  After this selection, no clear $\eta_c$ signal is
found in the distribution of $m_{\kp\kn\pip\pin\pip\pin}$, as shown in
Fig.~\ref{phi4pi}. Using Bayes method, a fit to $\eta_c\ar\etackk$
with a Breit-Wigner folded with a Gaussian and a polynomial background
gives 13.5 $\eta_c$ events at the 90\% confidence level.

From Monte-carlo simulation, in which
the angle ($\theta$) between the direction of the $e^+$ and $\etac$
in the laboratory frame is generated according to
a $1+\cos^2\theta$ distribution and uniform phase-space
is used for $\etac$ decaying into $\kp\kn 2(\pip\pin)$
and $\phi 2(\pip\pin)$, 
the detection efficiencies of
$\jpsi\ar\g\etac(\etac\ar\kp\kn\pip\pin\pip\pin)$, 
$\jpsi\ar\g\etac(\etac\ar K^{*0}\overline{K}^{*0}\pip\pin)$, and $\jpsi\ar\g\etac(
\etac\ar\phi \pip\pin\pip\pin)$ are determined as
$(1.43 \pm 0.04)\%$, $(1.36 \pm 0.04)\%$, and $(1.01\pm 0.02)\%$, respectively.
Therefore, the branching fractions obtained are
%$$B(\jpsi\ar\g\etac)\cdot B(\etac\ar K^+ K^-\pip\pin\pip\pin) =
%\frac{N_{obs}}{\eff \cdot N_{\jpsi}}=(1.21 \pm 0.32) \times 10^{-4}$$,

$B(\jpsi\ar\g\etac)\cdot B(\etac\ar K^+ K^-\pip\pin\pip\pin) =
(1.21 \pm 0.32) \times 10^{-4}$,

%$$B(\jpsi\ar\g\etac)\cdot B(\etac\ar K^{*0}\overline{K}^{*0} \pip\pin) <
%\frac{N_{obs}}{\eff \cdot N_{\jpsi}\cdot B(K^{*0}\ar\kp\pin)\cdot
%B(\overline{K}^{*0}\ar\kn\pip)}=1.86 \times 10^{-4}$$
$B(\jpsi\ar\g\etac)\cdot B(\etac\ar K^{*0}\overline{K}^{*0}
\pip\pin)= (1.29\pm0.43)\times 10^{-4}$ \\
$B(\jpsi\ar\g\etac)\cdot B(\etac\ar K^{*0}\overline{K}^{*0} \pip\pin) <1.86 \times 10^{-4}$,

and 

$B(\jpsi\ar\g\etac)\cdot B(\etac\ar\phi\pip\pin\pip\pin)< 4.72 \times 10^{-5}$.

\section{Analysis of $\jpsi\ar\g\etac$,$\etac\ar\pi^+\pi^-\pi^+\pi^-\pi^+\pi^-$}
\label{6pi}

These events are observed in the topology
$J/\psi \to \g\pi^+\pi^-\pi^+\pi^-\pi^+\pi^-$.
Events with six good charged tracks and at least one isolated photon
are selected. No particle identification is required.
To suppress background, a 4C kinematic fit is
performed under the hypothesis $\g\pi^+\pi^-\pi^+\pi^-\pi^+\pi^-$,
and the $\chi^{2}$ is required to be less than 10.
To reject background from $J/\psi\rightarrow{3(\pi^+\pi^-)}$ and 
$J/\psi\rightarrow{3(\pi^+\pi^-)}\pio$, we require 
$\chi^{2}_{\g\pi^+\pi^-\pi^+\pi^-\pi^+\pi^-}$ to be less than
$\chi^{2}_{\pi^+\pi^-\pi^+\pi^-\pi^+\pi^-}$ and
$\chi^{2}_{\pi^+\pi^-\pi^+\pi^-\pi^+\pi^-\pio}$.

\begin{figure}[htpb]
\centerline{\includegraphics[width=0.5\textwidth,height=0.32\textheight]
        {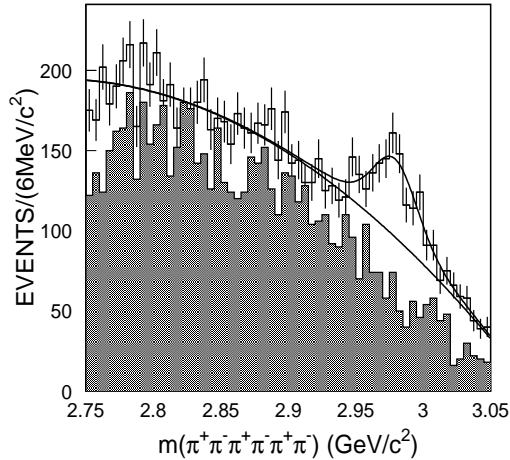}}
\caption{The distribution of $m_{\pi^+\pi^-\pi^+\pi^-\pi^+\pi^-}$ for
        selected events. 
The histogram with error bars
is data, the shaded part is the background estimated from Monte Carlo
simulation, and the curve is the fitting result described in the text.}
\label{etac1}
\end{figure}

Figure~\ref{etac1} shows the
${\pi^+\pi^-\pi^+\pi^-\pi^+\pi^-}$ invariant 
mass spectrum after the above selection. 
A clear $\etac$ peak is observed. The shaded histogram in Fig.~\ref{etac1} 
corresponds to background estimated from 58 million 
$\jpsi\ar anything$ Monte-Carlo events generated using the Lund-Charm 
generator~\cite{chenjc}, where no $\etac$ signal is evident.
A fit of the $m_{\pi^+\pi^-\pi^+\pi^-\pi^+\pi^-}$ distribution, 
which is shown as the solid curve in Fig.~\ref{etac1},
using a Breit-Wigner convoluted with a Gaussian to represent the
signal and a polynomial for the background, yields 
$427 \pm 64$ $\etac$ events with a statistical significance
of $6.9 \sigma$. In the fit, the mass and width of $\eta_c$ are
again fixed to PDG values~\cite{pdg2004}.

The detection efficiency for $\jpsi\ar\g\etac,
\etac\ar\pip\pin\pip\pin\pip\pin$ is determined to be
$(3.21 \pm 0.04)\%$, by Monte-Carlo simulation with 
the distribution of $\theta$, the angle between the directions of 
$e^+$ and $\etac$ in the laboratory frame, being generated with a $1+\cos^2\theta$ and with   $\etac$ decaying into 
$3(\pip\pin)$ being generated with 
a uniform phase-space distribution. The branching ratio is then found to be

%$$B(\jpsi\ar\g\etac)\cdot B(\etac\ar \pip \pin\pip\pin\pip\pin) =
%\frac{N_{obs}}{\eff \cdot N_{\jpsi}} =(2.59 \pm 0.32) \times 10^{-4}$$

$B(\jpsi\ar\g\etac)\cdot B(\etac\ar \pip \pin\pip\pin\pip\pin) =
 (2.59 \pm 0.32) \times 10^{-4}$.

\section{Systematic errors }
\label{syserreta}
The systematic errors mainly come from the following sources:

~~(1) MDC tracking efficiency \\
This has been measured with clean channels like
$\jpsi\ar\Lambda \bar{\Lambda}$ and
$\psi(2S)\ar\pip\pin\jpsi, \jpsi\ar\mu^+\mu^-$. It is found that the Monte
Carlo simulation agrees with data within 1-2\% for each charged track.
Therefore, 12\% is conservatively taken as the systematic error in 
the tracking efficiencies for the 6-prong final states analyzed here.

~~(2) Photon detection efficiency\\
This has been studied using different 
methods with $\jpsi\ar\rhoo\pio$ events~\cite{pheff}. The difference
between data and Monte Carlo simulation is less than 2\% for each
photon, and 2\% is taken as the systematic error for the photon
efficiency in
this analysis.

~~(3) Particle identification (PID)\\
This has been studied with $\jpsi\ar\kp\kn\pio$.
The efficiency of the PID from data is consistent with that from 
Monte Carlo simulation. The average difference is less than 2\%. For 
$\jpsi\ar\g\kp\kn\pip\pin\pip\pin$ decay, 4\% is taken as the systematic
error from PID. 

~~(4) Kinematic Fit \\
The kinematic fit is useful to reduce background. Using the same method
for estimating the systematic error as in Ref.~\cite{rhopi}, the decay mode 
 $\jpsi\ar 3(\pip\pin)\pio$ is also
analyzed. The efficiency difference of the kinematic fit for data and
Monte Carlo is 7.7\%. 
Since the decay of $\jpsi\ar 3(\pip\pin)\pio$ is similar to the two
channels analyzed in this paper, 7.7\% is also taken here as the
systematic error of the kinematic fit.

~~(5) $\eta_c$ parameters \\
Although the $\etac$ signal is clear, the number of events is not
large enough to determine the Breit-Wigner parameters and the
background shape well. The variation of the fit solution due to
changes of the $\eta_c$ mass and width corresponding to the
uncertainties in the PDG, as well as changes in the fitting mass
region used, is taken as a systematic error and listed in Table
\ref{syserr}.

~~(6) Background \\
For $\etac\ar\etackk$, the biggest background comes from
$\etac\ar\ks\ks\kp\kn$. When the
invariant mass of $\pip\pin$ is required to be within the $\ks$ mass
region ($|m_{\pip\pin}-0.497|<0.02$ GeV/c$^{2}$), five events remain in
the $\etac$ mass region. If all of them are regarded as signal
from $\etac\ar\ks\ks\kp\kn$, the background from this
decay mode is about 5.1\%, and this is taken as the systematic error
associated with background for this channel. No events remain for
$\etac\ar K^{*0}\overline{K}^{*0}\pip\pin$ and the upper limit is 
2.3 events at 90\% confidence level. Then the uncertainty caused by
$\etac\ar\ks\ks\pip\pin$ is 5.1\%.

For the $\etac\ar 3(\pip\pin)$, Monte Carlo simulation is used to
estimate the background from $\etac\ar \ks\ks\pip\pin$. Using the
branching fraction for $\etac\ar K^{0}\overline{K}^{0}\pip\pin$, obtained
from $B(\etac\ar \kp\kn\pip\pin)$~\cite{pdg2004}, Monte Carlo
simulation indicates that 33 background events contribute to the
$\etac$ signal. Compared to the 416 signal events from fitting the mass
spectrum, the background fraction is 7.9\% which is taken as the
background systematic error for this channel.

~~(7) Number of $\jpsi$ events\\ 
The number of $\jpsi$ events is $(57.70\pm 2.72)\times 10^6$, 
determined from $\jpsi$ inclusive four-prong events. 
The uncertainty is
taken as a systematic error in the branching ratio measurement. 
Table \ref{syserr} lists the systematic errors from all sources, and 
the total systematic error is the sum of them added in quadrature.

\begin{table}[htpb]
\caption{Systematic error sources and contributions (\%)}
\begin{center}
\begin{tabular}{ l|c|c|c|c}
\hline
 Sources & $\kp\kn 2(\pip\pin)$&$K^{*0}\overline{K}^{*0}\pip\pin$ &$\phi
2(\pip\pin)$
&$3(\pi^+\pi^-)$ \\
\hline
MDC tracking       & 12    & 12       &   12  & 12\\
Paticle ID         & 4     & 4        & 4     & negligible\\
Photon  efficiency & 2     &  2      & 2     & 2\\
Kinematic fit      & 7.7   & 7.7      & 7.7   & 7.7\\
$\etac$ parameters &9.9    & 18.6     & 14.7  & 7.4\\
MC statistics      & 2.6   & 2.9      & 2.9   & 1.1\\
Background uncertainty         & 5.1   &  5.1     &    & 7.9\\
$B(\phi\ar\kp\kn)$ &       &         &   1.4    & \\
Number of $\jpsi$ events & 4.7 & 4.7 & 4.7 & 4.7\\
\hline
Total  &19.4 & 25.0 & 21.7 &18.6\\
\hline
\end{tabular}
\label{syserr}
\end{center}
\end{table}

%%%%%%%%

\section{Results}
The decays of $\etac\ar\etackk$ and $\etac\ar\pip\pin\pip\pin\pip\pin$
are observed for the first time, and their decay branching ratios are
measured. The upper limits of $\etac\ar\phi\pip\pin\pip\pin$ and
$\etac\ar K^{*0}\overline{K}^{*0}\pip\pin$ are also set at the 90\%
confidence level. To conservatively estimate the upper limit, the
systematic error is included by lowering the efficiency by one standard
deviation.  Table \ref{result} shows the branching ratio results
including systematic errors.

\begin{table}[htpb]
\caption{Numbers used in the calculations of branching fractions and upper
limits.}
\begin{center}
\begin{tabular}{c|c|c|c}
\hline
Decay Modes   & $N_{obs}$ & $\eff$ (\%)  & Branching Fraction \\
\hline
$\jpsi\ar\g\etac, \etac\ar \kp\kn2(\pip\pin)$ & 100$\pm$26  & 1.43$\pm$0.04  &
$(1.21\pm 0.32 \pm 0.23)\times 10^{-4}$   \\
\hline
$\jpsi\ar\g\etac, \etac\ar  K^{*0}\overline{K}^{*0}\pip\pin$ & $45\pm15$  &
$1.36\pm 0.04$ & $(1.29\pm 0.43\pm 0.32)\times 10^{-4}$   \\

$\jpsi\ar\g\etac, \etac\ar  K^{*0}\overline{K}^{*0}\pip\pin$ & $<65$  & 
$1.36\pm 0.04$
& $<2.46\times 10^{-4}$ (90\% C.L.)  \\
\hline
$\jpsi\ar\g\etac, \etac\ar \phi2(\pip\pin)$ & $<13.5$  & $1.01\pm 0.02$
& $<6.03\times 10^{-5}$ (90\% C.L.)  \\
\hline
$\jpsi\ar\g\etac,\etac\ar 3(\pip\pin)$ & 427$\pm$64  & 3.21$\pm$0.04    &
$(2.59\pm 0.32 \pm 0.48)\times 10^{-4}$   \\

\hline
\end{tabular}
\end{center}
\label{result}
\end{table}

Using the branching fraction of $\jpsi\ar\g\etac$ as 
$B(\jpsi\ar\g\etac)=(1.3\pm 0.4)\%$ from the PDG~\cite{pdg2004}, we obtain:
\begin{center}
$ B(\etac\ar\etackk)= (0.93\pm 0.25 \pm 0.34)\times 10^{-2}$
\end{center}

\begin{center}
$B({\etac}\rightarrow K^{*0}\overline{K}^{*0}\pip\pin)
= (0.99\pm 0.33\pm0.39)\times 10^{-2}$
\end{center}

\begin{center}
$B({\etac}\rightarrow K^{*0}\overline{K}^{*0}\pip\pin)
<2.36\times 10^{-2}$
\end{center}

\begin{center}
$B(\etac\ar\phi\pip\pin\pip\pin)< 5.81 \times
10^{-3}$  (90\% C.L.)
\end{center}

\begin{center}
$B({\etac}\rightarrow{\pi^+\pi^-\pi^+\pi^-\pi^+\pi^-})
= (1.99\pm0.25\pm0.72)\times10^{-2}$
\end{center}

%\acknowledgments

   The BES collaboration thanks the staff of BEPC and the computing 
center for their hard efforts.
This work is supported in part by the National Natural Science Foundation
of China under contracts Nos. 19991480, 10225524, 10225525, the Chinese Academy
of Sciences under contract No. KJ 95T-03, the 100 Talents Program of CAS
under Contract Nos. U-11, U-24, U-25, and the Knowledge Innovation Project
of CAS under Contract Nos. U-602, U-34 (IHEP); and by the
National Natural Science Foundation of China under Contract
No.10175060 (USTC), and No. 10225522 (Tsinghua University); and by the
U. S. Department of Energy under Contract N0. DE-FG02-04ER41291.

\end{document}